\documentclass[figure]{pasj00}
\tenpoint
\SetRunningHead{Astronomical Society of Japan}{
            Gravitational Instability of Shocked Interstellar Gas Layers}

\title{Gravitational Instability of Shocked Interstellar Gas Layers}
\author{Kazunari \textsc{Iwasaki} and  Toru \textsc{Tsuribe}}
\affil{%
   Department of Earth and Space Science, Osaka University, Machikaneyama 1-1 
   Toyonaka 560-0043, Osaka}
\email{iwasaki@vega.ess.sci.osaka-u.ac.jp, tsuribe@vega.ess.sci.osaka-u.ac.jp}
\KeyWords{{Hydrodynamics} --- {Instability} --- {Interstellar: clouds} --- 
{Stars: formation}}
\Received{$\langle$reception date$\rangle$}
\Accepted{$\langle$acception date$\rangle$}
\Published{$\langle$publication date$\rangle$}
\begin{document}
\maketitle

%%%%%%%%%%%%%%%%%%%%%%%%%%%%%%%%%%%%%%%%%%%%%%%%%%%%%%%%%%%%%%%%%%%%%%%%%%%%%%
\begin{abstract}
In this paper we investigate gravitational instability of shocked gas layers 
using linear analysis. An unperturbed state is a self-gravitating isothermal layer which 
grows with time by the accretion of gas through shock fronts due to a cloud-cloud collision.
Since the unperturbed state is not static, and cannot be described by 
a self-similar solution, we numerically solved the perturbation equations and directly 
integrated them over time.  We took account of the distribution of physical 
quantities across the thickness.
Linearized Rankine-Hugoniot relations were imposed at shock fronts 
as boundary conditions.  
The following results are found from our unsteady linear analysis: 
the perturbation initially evolves in 
oscillatory mode, and begins to grow at a certain epoch. 
The wavenumber of the fastest growing mode is given 
by $k=2\sqrt{2\pi G\rho_\mathrm{E} {\cal M\mit}}/c_\mathrm{s}$, 
where $\rho_\mathrm{E},\;c_\mathrm{s}$ and $\cal M\mit$ are 
the density of parent clouds, the sound velocity and 
the Mach number of the collision velocity, respectively. 
For this mode, the transition epoch from oscillatory to growing mode 
is given by $t_g = 1.2/\sqrt{2\pi G\rho_\mathrm{E} {\cal M\mit}}$. 
The epoch at which the fastest growing mode becomes non-linear is given by
$2.4\delta_0^{-0.1}/\sqrt{2\pi G \rho_\mathrm{E}{\cal M\mit}}$, where $\delta_0$ is 
the initial amplitude of the perturbation of the column density.
As an application of our linear analysis, we investigate criteria for 
collision-induced fragmentation.
Collision-induced fragmentation will occur
only when parent clouds are cold, or $\alpha_0=5c_\mathrm{s}^2 R/2G M < 1$,
where $R$ and $M$ are the radius and the mass of parent clouds, respectively.
\end{abstract}
%%%%%%%%%%%%%%%%%%%%%%%%%%%%%%%%%%%%%%%%%%%%%%%%%%%%%%%%%%%%%%%%%%%%%%%%%%%%%%
\section{Introduction}
%%%%%%%%%%%%%%%%%%%%%%%%%%%%%%%%%%%%%%%%%%%%%%%%%%%%%%%%%%%%%%%%%%%%%%%%%%%%%%
Shock waves that propagate in interstellar gases play important roles 
in the formation of stars and other structures in the universe. 
Shock waves are formed by 
cloud-cloud collisions (\cite{H94}, \cite{S00}), supernova explosions (\cite{HA77}),
expansions of HII regions by ionizing photons emitted from OB stars (\cite{B64}),
 and so on. Shock waves sweep out external gases and form highly compressed 
sheet-like structures behind the shock fronts. These layers, which are sheet-like
structures, are often supposed to fragment and eventually to form stars.

Shock compression contributes especially on the formation of aggregations of stars.
\citet{EL77} proposed a scenario of a sequential formation of OB star subgroups.
The expansion of a HII region that is formed by an OB subgroup sweeps up the 
gases. A highly compressed layer forms, and then possibly fragments.
Thereafter, a new OB star subgroup forms from the layer,
and the ionizing photons emitted from 
the new subgroup induce the formation of next subgroup by the same process.
\citet{KBF93} proposed a scenario of a globular cluster formation 
that is induced by the gravitational instability of a layer formed by 
a supersonic collision between large proto-grobular cluster clouds. 
Observations showed that galaxies containing young star clusters have larger
random velocity dispersions of gases than those that do not have young clusters
(\cite{FN90}).
If the large random velocity dispersion of 
gases is stimulated by mergers or interactions between galaxies (\cite{AZ92}),
the scenario of star cluster formation
by cloud-cloud collision is indicated.

In the process of star cluster formation from the layer, it is crucial to understand
when and how a shock-compressed layer fragments. 
\citet{GL65} (hereafter GL65) and \citet{EE78} 
investigated the stability of a static 
isothermal self-gravitating layer that is supported by constant external 
pressure on both sides using linear analysis.
They showed that the wave length of the most unstable mode
is approximately the scale height of gravity, 
$H_0=c_\mathrm{s}/\sqrt{2\pi G \rho_{00}}$ 
when the density at the central plane, $\rho_{00}$, is larger than the 
boundaries, $\rho_\mathrm{b}$, of the layer ($\rho_{00}/\rho_\mathrm{b}>1$). 
When $\rho_{00}/\rho_\mathrm{b}\simeq 1$, the wave length of the most unstable mode is
the thickness of the layer. 
The maximum growth rate is about the free-fall growth rate of the layer in both cases.

However, a layer formed by shock compression is bounded by shock waves on at
least one side. The effects of the shock boundary strongly influence
the stability of the layer when $\rho_{00}/\rho_\mathrm{b}\simeq 1$. 
This is because the layer is mainly confined not by self-gravity,
but by external pressure. 
\citet{V94} (hereafter V94) derived 
a simple dispersion relation of a layer bounded by shocks on both sides 
using shock boundary conditions, while
neglecting the time dependence of the column density. 
The dispersion relation of the shocked layer is quite different from 
that of the static layer when 
$\rho_{00}/\rho_\mathrm{b}\simeq 1$ (see Appendix \ref{appe 1zone} and \ref{const}).
Therefore, the results with a constant pressure boundary
are not applicable to a layer bounded by a shock front
when $\rho_{00}/\rho_\mathrm{b}\simeq 1$.
Since the layer experiences the state of $\rho_{00}/\rho_\mathrm{b}\simeq 1$ 
in early phase from the moment when the layer is formed by collision, 
an analysis that takes account of the effect of the shock boundary is important. 

In the analysis of V94, the time evolution of 
the column density is not considered. In this paper, in order to consider the fastest 
growing mode in the layer that is formed by cloud-cloud collision, we take into 
account both shock boundary and time-dependent column density. 
The distribution of the physical variables across the thickness is also taken 
into account accurately. 

In section 2, we set up a model of a layer. 
Before our linear analysis,
previous works (V94; \cite{W94}) using a 1-zone model is reviewed.
In section 3, an unperturbed state is described. In section 4,
perturbation equations are derived and a numerical method, 
boundary conditions and initial conditions are shown. The results are shown in section 5. 
Criteria for collision-induced fragmentation
are discussed in section 6. Our study is summarized in section 7.
%%%%%%%%%%%%%%%%%%%%%%%%%%%%%%%%%%%%%%%%%%%%%%%%%%%%%%%%%%%%%%%%%%%%%%%%%%%%%%
\section{Model}
%%%%%%%%%%%%%%%%%%%%%%%%%%%%%%%%%%%%%%%%%%%%%%%%%%%%%%%%%%%%%%%%%%%%%%%%%%%%%%
%%%%%%%%%%%%%
\subsection{Assumptions}\label{assumption}
%%%%%%%%%%%%%
\begin{figure}
 \begin{center}
  \FigureFile(80mm,40mm){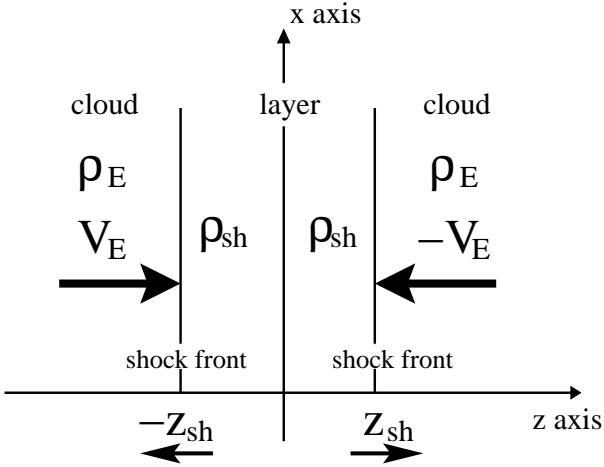}
 \end{center}
 \caption{Schematic picture of a compressed layer between the shock fronts.}
\label{shock}
\end{figure}
%%%
We consider a head-on collision of two physically identical clouds with 
sufficiently large Mach number (${\cal M\mit}\gg 1$).
The equation of state is assumed to be isothermal. 
In this paper, Cartesian coordinates $(x,y,z)$ are used and 
the $z$-axis is perpendicular to a layer. The layer is assumed to extend 
infinitely in the $x-y$ plane. We can discuss the system in the $x-z$ plane without 
any loss of generality in the linear regime.
Physically identical clouds are assumed to collide along the $z$-axis with 
velocity $\pm V_\mathrm{E}$ and density $\rho_\mathrm{E}$ 
at $t=0$ and $z=0$. The two shock fronts at $z=\pm z_\mathrm{sh}$  
propagate in both directions and 
a highly compressed layer is formed between two shock fronts 
(see figure \ref{shock}).
The isothermal shock boundary conditions are given by
%%%
\begin{equation}
\rho_{\mathrm{sh}}=\rho_\mathrm{E} {\cal M\mit}^2,\;\;
\dot{z}_{\mathrm{sh}}=\frac{c_\mathrm{s}}{\cal M\mit},\;\;\mathrm{and}\;\;
V_\mathrm{E} = \left({\cal M\mit} - \frac{1}{\cal M\mit}\right)c_\mathrm{s},
\label{1zone boundary}
\end{equation}
%%%
where $\cal M\mit=V_\mathrm{sh}/c_\mathrm{s}$ is the Mach number and
$V_\mathrm{sh}$ is the velocity of the clouds in the rest frame of the shock front
(see Appendix \ref{shock boundary} in detail).
The unperturbed column density, $\Sigma_0$, is defined by 
%%%
\begin{equation}
\Sigma_0(t) = \int_{-z_\mathrm{sh}}^{z_\mathrm{sh}}\rho_0(z) \mathrm{d}z
= 2\rho_\mathrm{E}{\cal M\mit}c_\mathrm{s}t,
\end{equation}
%%%
which increases as $\propto t$ by gas accretion. 
Hereafter, we investigate gravitational instability of the layer between two shock
fronts.
%%%%%%%%%%%%%%%
\subsection{Dispersion Relation under 1-zone Model}\label{1zone}
%%%%%%%%%%%%%%%
%%%
\begin{figure*}[t]
 \begin{center}
  \begin{tabular}{cc}
     \hspace{-3mm}
     \FigureFile(80mm,30mm){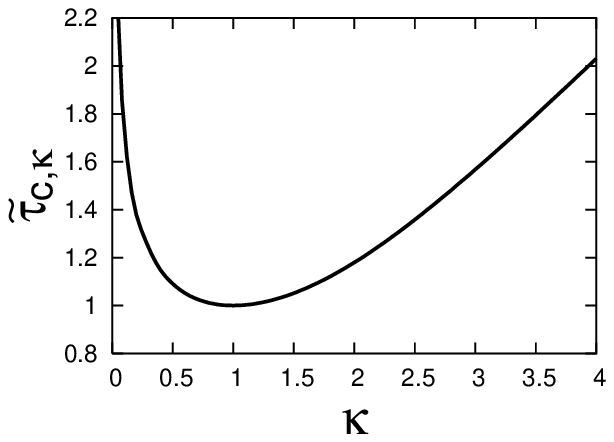} 
     \hspace{-6mm}
     &
     \FigureFile(80mm,30mm){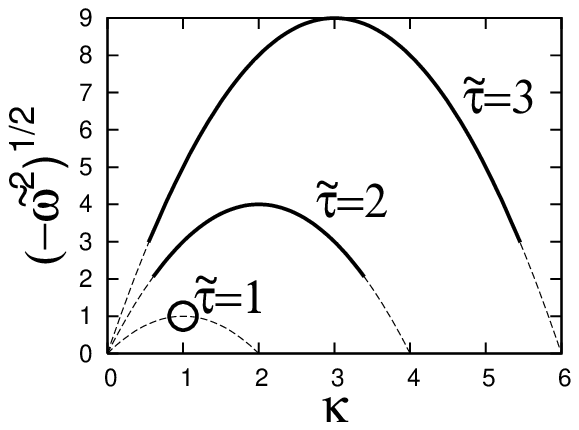} \\
     {\scriptsize (a)}
     \hspace{-6mm}
     &
     {\scriptsize (b)} \\
  \end{tabular}
 \end{center}
  \caption{(a)Dependence of the transition epoch, $\tilde{\tau}_{\mathrm{c},\kappa}$,
           from oscillatory to growing mode 
           on wavenumbers 
           $\kappa$ in the 1-zone model. The curve shows a plot of equation (\ref{tau c}).
           (b)Dispersion relation [equation (\ref{disp 1zone})] at 
           $\tilde{\tau}$=1, 2 and 3, where $\tilde{\omega}=t_\mathrm{c}\omega$. 
           The solid lines satisfy $\sqrt{-\omega^2}>1/t$, where perturbation
           grows faster than growth of the column density.
           The open circle indicates the mode that begins to grow at the earliest 
           epoch, $\kappa=1$.
           }\label{begin tau}
\end{figure*}
As an essential guide for our linear analysis,
a simple dispersion relation using a 1-zone model is summarized (see V94 in detail).
In the 1-zone model, we consider averaged physical quantities across the thickness. 
The layer is assumed to be geometrically thin, i.e., $z_{\mathrm{sh}}/H_0\ll1$
and $kL_0 \ll 1$,
where $L_0$ and $H_0=c_\mathrm{s}/\sqrt{2\pi G \rho_{\mathrm{sh}}}$ are 
the thickness and the scale height of the layer, respectively.
Averaging physical variables across the thickness and neglecting 
the time evolution of $\Sigma_0(t)$ with shock boundary condition, V94 derived 
an approximate dispersion relation which is given by
%%%
\begin{equation}
\omega^2 \sim c_\mathrm{s}^2 k^2 - 2\pi G k \Sigma_0(t).
\label{disp 1zone}
\end{equation}
%%%
The derivation of equation (\ref{disp 1zone}) is shown in Appendix \ref{appe 1zone}.
The dispersion relation is identical to that of an infinitely thin disk 
without boundary condition. 
In Appendix \ref{const}, we derive a dispersion relation of a static layer which 
is confined by constant external pressure by the using 1-zone model for comparison.

According to equation (\ref{disp 1zone}), the time evolution of the perturbations
with wavenumber $k$ is described as follows: 
in the early phase $t\sim 0$, the equation (\ref{disp 1zone}) is approximately
$\omega^2 \sim c_\mathrm{s}^2k^2$, because $\Sigma_0(t)$ is small.
Thus, the perturbations are initially in an oscillatory mode.
Since $c_\mathrm{s}^2 k^2$ is constant with time, while $2\pi G k\Sigma_0(t) $
increases with time, the perturbations 
will change from the oscillatory mode to a growing mode at a certain epoch.

This transition epoch can be derived by a simple order estimation.
The gravitational instability sets in if $\sqrt{-\omega^2}>1/t$
(\cite{W94}), where
$1/t$ is the evolution rate of the unperturbed layer by gas accretion.
The transition epoch, $t_{\mathrm{c},k}$, at which perturbations 
with wavenumber $k$ begin to grow, is given by
%%%
\begin{equation}
\frac{1}{t_{\mathrm{c},k}}=\sqrt{-\omega^2}=
\sqrt{2\pi G k\Sigma_0(t_{\mathrm{c},k}) - c_\mathrm{s}^2k^2}.
\label{tau c1}
\end{equation}
%%%
Using the characteristic timescale and the wavenumber defined by
%%%
\begin{equation}
t_{\mathrm{c}}=1/\sqrt{2\pi G \rho_\mathrm{E}{\cal M\mit}} \;\;\;\mathrm{and}\;\;\;
k_{\mathrm{c}}=\sqrt{2\pi G\rho_\mathrm{E}{\cal M\mit}}/c_\mathrm{s},
\label{tau kappa def}
\end{equation}
%%%
equation (\ref{tau c1}) is rewritten as an independent form on Mach number as
%%%
\begin{equation}
\frac{1}{\tilde{\tau}_{\mathrm{c,\kappa}}}
=\sqrt{2\tilde{\tau}_{\mathrm{c},\kappa}\kappa - \kappa^2},
\label{tau c}
\end{equation}
%%%
where $\tilde{\tau}=t/t_\mathrm{c}$ and $\kappa=k/k_\mathrm{c}$.
Figure \ref{begin tau}a shows the dependence of $\tilde{\tau}_{c,\kappa}$ on 
the wavenumber $\kappa$.
Figure \ref{begin tau}a indicates that the mode with $\kappa=1$ 
begins to grow at the earliest epoch $\tilde{\tau}=1$.
Then, the perturbations with $\kappa\sim 1$ 
switch to the gravitational instability successively. 
\citet{W94} estimated fragmentation mass scale using $\kappa=1$. 
However, the earliest transition mode, $\kappa=1$, is not 
the fastest growing mode in later epoch. 
Figure \ref{begin tau}b represents the dispersion relation 
[equation (\ref{disp 1zone})]
at $\tilde{\tau}$=1, 2 and 3, where $\tilde{\omega}=t_\mathrm{c}\omega$.
The open circle indicates the earliest transition mode, $\kappa=1$.
The gravitationally unstable state that satisfies $\sqrt{-\omega^2}>1/t$ 
is shown by the solid lines at each epoch. 
Figure \ref{begin tau}b indicates that $\tilde{\omega}$ and $\kappa$ of 
the fastest growing mode 
varies with time. 

In the discussions so far, have been based on equation (\ref{disp 1zone}),
which was derived by neglecting the time dependence of $\Sigma_0(t)$. 
However, especially during $t \ltsim t_{\mathrm{c},k}$,  
the evolution rate of the layer $1/t$
is larger than $\sqrt{-\omega^2}$. Therefore,
the evolution of $\Sigma_0(t)$ can not be neglected. 
In this paper, in order to accurately predict the fragmentation scale of the evolving layer at the 
earliest epoch, the evolution of the unperturbed layer is
taken into account consistently. 
In the following sections, we investigate the gravitational 
instability in the linear regime with a time-dependent $\Sigma_0(t)$, as well as
shock boundary conditions without the 1-zone approximation.
%%%%%%%%%%%%%%%%%%%%%%%%%%%%%%%%%%%%%%%%%%%%%%%%%%%%%%%%%%%%%%%%%%%%%%%%%%%%%%
\section{Unperturbed State}\label{unperturbed state}
%%%%%%%%%%%%%%%%%%%%%%%%%%%%%%%%%%%%%%%%%%%%%%%%%%%%%%%%%%%%%%%%%%%%%%%%%%%%%%
In our linear analysis, the unperturbed state is the time-evolving layer 
formed by cloud collision. 
To begin with, we define the distribution of the density, $\rho(z,t)$.
The density distribution of a hydrostatic isothermal self-gravitating layer
that is supported by external pressure, $P_\mathrm{b}$, is analytically written as 
%%%
\begin{equation}
\rho(z) = \frac{\rho_{00}}{\cosh^2(z/H_0)},\;\;
H_0 = \frac{c_\mathrm{s}}{\sqrt{2\pi G \rho_{00}}}.
\label{equilibrium}
\end{equation}
%%%
The relationship among $\rho_{00}$, $P_b$ and $\Sigma_0$ is given by
%%%
\begin{equation}
\rho_{00} = \frac{P_{\mathrm{b}}}{c_\mathrm{s}^2} + \frac{\pi G \Sigma_0^2}{2c_\mathrm{s}^2}.
\label{den00 Pb column}
\end{equation}
%%%
Equation (\ref{den00 Pb column}) shows that a set of external pressure
$P_{\mathrm{b}}$ and the column density $\Sigma_0$ provides
a unique density distribution $\rho(z)$. 

The sound-crossing time across the thickness $L_0$ and evolution time are defined by 
$t_\mathrm{cross}=L_0/c_\mathrm{s}$ and 
$t_\mathrm{evo}=\Sigma_0/(\mathrm{d}\Sigma/\mathrm{d}t)=t$, respectively. 
Since the thickness $L_0$ varies as $\ltsim 2c_\mathrm{s} t/{\cal M\mit}$,
the ratio between the two timescales is 
$t_{\mathrm{cross}}/t_\mathrm{evo} \lesssim 2/{\cal M\mit}\ll 1$.
Therefore, we assume that the layer evolves while maintaining the equilibrium density 
distribution in the $z$-direction with a large Mach number.

Next, we assume that the Mach number, ${\cal M\mit}$ at the shock front 
is constant. 
We consider the evolution of the shock front in the rest frame of the $z=0$ plane. 
According to equation (\ref{1zone boundary}), if $V_\mathrm{E}$ is constant, 
the shock front velocity, $\dot{z}_{\mathrm{sh}}$, is constant without gravity.
If gravity is taken into account, as the effect of gravity becomes important, 
$\dot{z}_\mathrm{sh}$ decreases with time. 
Consequently, in the rest frame of shock front, the velocity of accreting gas, 
$V_\mathrm{sh}=V_\mathrm{E}+\dot{z}_\mathrm{sh}$, 
and the Mach number ${\cal M\mit}$ decrease.
However the fractional change of Mach number is as small as
${\cal M\mit}^{-2}\ll1$.
Therefore, the constant $\cal M\mit$ assumption well approximates a
constant $V_\mathrm{E}$.

The velocity distribution in the postshock layer is assumed to be zero.
In reality, $v_z(z,t)$ is not zero in order to
attain the hydrostatic state at each instant of time. 
However, this assumption
is also valid when the Mach number is sufficiently large. The validity of 
this assumption is discussed in Appendix \ref{valid}.

The above assumptions completely provide the time evolution of the density distribution  
and the position of the shock front analytically.
With a constant Mach number, 
the column density and the external pressure are given by  
%%%
\begin{equation}
\Sigma_0(t) = 2\rho_\mathrm{E} {\cal M\mit}c_\mathrm{s}t,\hspace{5mm}
P_{\mathrm{b}} = c_\mathrm{s}^2\rho_\mathrm{E} {\cal M\mit}^2.
\label{column den}
\end{equation}
%%%
Substituting equation (\ref{column den}) into equation (\ref{den00 Pb column}), 
the density at the center is obtained as
%%%
\begin{equation}
\rho_{00} = \rho_\mathrm{E} {\cal M\mit}^2(1+\tau^2),
\label{den00}
\end{equation}
%%%
where $\tau$ is defined by $\tau = t\sqrt{2\pi G \rho_\mathrm{E}}$, which is 
the time normalized by the free-fall timescale in the preshock region.
Equation (\ref{den00}) shows that the layer has approximately uniform density 
distribution, $\rho_{00}/\rho_\mathrm{b}\sim 1$, at $\tau\ll 1$, 
and $\rho_{00}/\rho_\mathrm{b}=2$ at $\tau = 1$.
Because the density at the shock front is given by $\rho_\mathrm{E}{\cal M\mit}^2$,
the time evolution of $z_\mathrm{sh}(t)$ is written with 
equation (\ref{equilibrium}) as
%%%
\begin{eqnarray}
z_{\mathrm{sh}} &=& H_0\cosh^{-1}\sqrt{1+\tau^2} \nonumber \\
&=&\frac{c_\mathrm{s}}{\cal M\mit}
\frac{1}{\sqrt{2\pi G\rho_\mathrm{E}}}\frac{\cosh^{-1}\sqrt{1+\tau^2}}{\sqrt{1+\tau^2}}\;.
\label{zsh}
\end{eqnarray}
%%%
When $\tau \ll 1$, equation (\ref{zsh}) becomes 
$z_{\mathrm{sh}}\sim c_\mathrm{s} t/{\cal M\mit}$. 
As $\tau$ increases, the increasing rate of $z_\mathrm{sh}$ is suppressed by self-gravity.
At $\tau \sim 1.5$, $z_\mathrm{sh}$ reaches the maximum
and begins to decrease.
%%%%%%%%%%%%%%%%%%%%%%%%%%%%%%%%%%%%%%%%%%%%%%%%%%%%%%%%%%%%%%%%%%%%%%%%%
\section{Perturbed State}
%%%%%%%%%%%%%%%%%%%%%%%%%%%%%%%%%%%%%%%%%%%%%%%%%%%%%%%%%%%%%%%%%%%%%%%%%
In this section, we consider perturbations on the unperturbed state, which is
defined in the previous section. 
To begin, the basic equations of ideal self-gravitating isothermal fluid 
are the mass conservation,
%%%
\begin{equation}
\frac{\partial \rho}{\partial t} + \boldsymbol{\nabla}
\cdot (\rho\boldsymbol{v})=0,
\label{mass conserv}
\end{equation}
%%%
the momentum conservation,
%%%
\begin{equation}
\frac{\partial}{\partial t}(\rho\boldsymbol{v})
+ \boldsymbol{\nabla}\cdot (\rho \boldsymbol{v}\otimes\boldsymbol{v} +
P \boldsymbol{I}) + \rho \boldsymbol{\nabla}\phi=0,
\label{momentum conserv}
\end{equation}
%%%
the Poisson equation, 
%%%
\begin{equation}
\boldsymbol{\nabla}^2\phi = 4\pi G \rho,
\label{poisson eq}
\end{equation}
%%%
and isothermal condition,
%%%
\begin{equation}
\frac{P}{\rho}=c_\mathrm{s}^2=\mathrm{const},
\end{equation}
%%%
where $\boldsymbol{I}$ is the unit matrix and $\phi$ is the gravitational potential.
The perturbation variables ($\delta \rho$, $\delta v_z$, 
$\delta v_x$ and $\delta \phi$) are defined by 
%%%
\begin{equation}
\rho(x,z,t) = \rho_0(z,t) + \delta \rho(z,t)e^{ikx}, 
\end{equation}
%%%
\begin{equation}
v_z(x,z,t)  = v_{z,0}(z,t) + \delta v_z(z,t)e^{ikx}, 
\end{equation}
%%%
\begin{equation}
v_x(x,z,t) = \delta v_x(z,t)e^{ikx}\;\;, 
\end{equation}
%%%
and 
%%%
\begin{equation}
\phi(x,z,t) =\phi_0(z,t) + \delta \phi(z,t)e^{ikx},
\end{equation}
%%%
where we take the Fourier mode in the $x$-direction.
Subscript "0" denotes quantities of the unperturbed state.
By linearizing the basic equations about the perturbation variables, 
the perturbation equations are given by
%%%
\begin{equation}
\frac{\partial \delta \rho}{\partial t}
+ \frac{\partial}{\partial z}(\rho_0 \delta v_z)
= -ik \rho_0 \delta v_x,
\end{equation}
%%%
\begin{equation}
\frac{\partial}{\partial t}(\rho_0\delta v_z)
+ \frac{\partial}{\partial z}\left(c_\mathrm{s}^2\delta \rho\right)  
=  - \delta \rho \frac{\partial \phi_0}{\partial z}
- \rho_0 \frac{\partial \delta \phi}{\partial z},
\end{equation}
%%%
\begin{equation}
\frac{\partial}{\partial t}(\rho_0\delta v_x)= -ik c_\mathrm{s}^2 
\delta \rho - ik\rho_0 \delta \phi,
\end{equation}
%%%
and 
%%%
\begin{equation}
\frac{\partial^2 \delta \phi}{\partial z^2} - k^2 \delta \phi= 4\pi G \delta \rho,
\end{equation}
%%%
where we assume $v_{z,0}=0$.

The rippling of the shock front $\delta z_\mathrm{sh}$ is defined by 
%%%
\begin{equation}
z_\mathrm{sh}(x,t) = z_\mathrm{sh,0}(t) + \delta z_\mathrm{sh}(t)e^{ikx}.
\end{equation}
%%%
The column density perturbation, which is defined by 
%%%
\begin{equation}
\delta \Sigma=
\int_{-z_\mathrm{sh}}^{z_\mathrm{sh}}\rho(x,z,t)\mathrm{d}z
- \Sigma_0,
\end{equation}
%%%
can be divided into two contributions, as follows:
%%%
\begin{equation}
\delta \Sigma=\rho_\mathrm{sh}\delta L + \delta \sigma,
\label{sigma two contri}
\end{equation}
%%%
where $\delta L = 2\delta z_\mathrm{sh}$ and 
$\delta \sigma=\int_{-z_\mathrm{sh,0}}^{z_\mathrm{sh,0}}\delta \rho\mathrm{d}z$.
The first term describes the fluctuation of the shock boundary, and the second term describes
the contribution of density perturbation inside the layer.

Since the unperturbed state is neither static nor self-similar solution, 
the perturbation equations are not reduced to some eigenvalue problems. 
Therefore, we directly integrate the perturbation equations over both time and 
the $z$-direction to investigate the time evolution of the perturbations.
%%%%%%%%%%%%%%%%%
\subsection{Numerical Method}
%%%%%%%%%%%%%%%%%
The upwind finite-difference method is used as numerical scheme.
Because the perturbation equations have source terms, 
the numerical flux is modified appropriately (\cite{HG00}).

We performed three tests of our code.
First, we calculated $k=0$ perturbation without self-gravity.
The results of our calculation were compared to the results of 
a one-dimensional, non-linear hydrodynamical calculation using 2nd-order 
Godunov code in the 
same situation. Our linear results agree with 
the non-linear calculation very well. Secondly, we calculated the case with $k\neq0$ 
without self-gravity.
V94 derived the dispersion relation of the even mode
to be $\omega =\pm c_\mathrm{s} k(1-2/{\cal M\mit}^2)$.
We obtained the same frequency as V94.
Thirdly, we calculated stability of the static self-gravitating layer. 
Our results agree with the growth rate of GL65.
In these test calculations, our code provided correct results.
%%%%%%%%%%%
\subsection{Boundary Conditions}
%%%%%%%%%%%
The boundary conditions are set at $z=0$ and $z=z_\mathrm{sh,0}$.
We investigate the even mode, and
the boundary conditions at $z=0$ are given by
%%%
\begin{equation}
\left(\frac{\partial \delta \rho}{\partial z}\right)_{z=0}=0,\;\;
\left(\frac{\partial \delta \phi}{\partial z}\right)_{z=0}=0,\;\;
\mathrm{and}\;\;
\delta v_z(z=0) = 0.
\end{equation}
%%%

At the shock front $z_\mathrm{sh,0}$, the physical variables must
satisfy the Rankine-Hugoniot relations. Thus, we impose linearized
Rankine-Hugoniot relations at $z_\mathrm{sh,0}$ (see Appendix \ref{shock boundary}
in detail), which are given by
%%%
\begin{equation}
\delta \rho_{\mathrm{sh}} = - \delta z_{\mathrm{sh}}
\left(\frac{\partial \rho_0}{\partial z}\right)_{z=z_{\mathrm{sh,0}}}
+ 2\rho_\mathrm{E}\frac{\cal M\mit}{c_\mathrm{s}}
\frac{\mathrm{d}\delta z_{\mathrm{sh}}}{\mathrm{d} t},
\end{equation}
%%%
\begin{equation}
\delta v_{z,\mathrm{sh}} = - \delta z_{\mathrm{sh}}
\left(\frac{\partial v_z}{\partial z}\right)_{z=z_{\mathrm{sh,0}}}
+ \left(1+\frac{1}{{\cal M\mit}^2}\right)
\frac{\mathrm{d}\delta z_{\mathrm{sh}}}{\mathrm{d} t},
\end{equation}
%%%
\begin{equation}
\delta v_{x,\mathrm{sh}} = -ik V_\mathrm{E} \delta z_{\mathrm{sh}}
\end{equation}
%%%
and 
%%%
\begin{equation}
\frac{\partial \delta \phi_\mathrm{sh}}{\partial z}
+ k\delta \phi_\mathrm{sh} + 
4\pi G (\rho_{\mathrm{sh}}-\rho_\mathrm{E})\delta z_{\mathrm{sh}}=0,
\end{equation}
where subscript "sh" indicates the value at the shock front.
%%%%%%%%%%%%%%%%%
\begin{figure}[h]
 \begin{center}
  \FigureFile(80mm,30mm){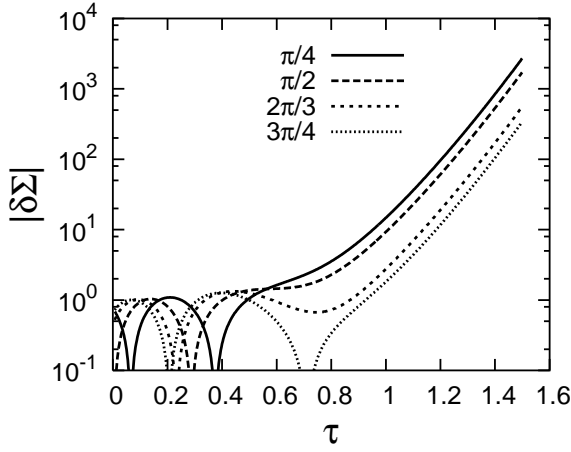}
 \end{center}
 \caption{Time evolution of $\delta \Sigma$ for $\kappa=4$ and ${\cal M\mit}=10$.
          Each line corresponds to initial phases  
          $\psi=\pi/4$, $\pi/2$, $2\pi/3$, $3\pi/4$, respectively.
          }\label{phase}
\end{figure}
\subsection{Initial Conditions}
%%%%%%%%%%%%%%%%%
The initial epoch is set to be $\tau_\mathrm{ini}=5\times10^{-3}$, but 
the results are not 
sensitive to $\tau_\mathrm{ini}$ as long as it is sufficiently small.
As an initial condition, an oscillatory mode (see subsection \ref{1zone}) 
without self-gravity is adopted.
The time evolution of perturbations depends on an initial phase $\psi$ of 
perturbation in oscillatory mode.
In figure \ref{phase}, the evolution of $\delta \Sigma$ with different
initial phases is shown for $\cal M\mit$=10 and $\kappa=4$. 
In this paper, we focus on the evolution with the 
initial phase by which $\delta \Sigma$ grows fastest.
%%%
%%%%%%%%%%%%%%%%%%%%%%%%%%%%%%%%%%%%%%%%%%%%%%%%%%%%%%%%%%%%%%%%%%%%%%%%%%
\section{Results}
%%%%%%%%%%%%%%%%%%%%%%%%%%%%%%%%%%%%%%%%%%%%%%%%%%%%%%%%%%%%%%%%%%%%%%%%%%
%%%%%%%%%%%%%%%%%
\subsection{The Evolution of Perturbations }
\label{dependence}
%%%%%%%%%%%%%%%%%
%%%
\begin{figure*}[t]
 \begin{center}
  \begin{tabular}{cc}
     \FigureFile(80mm,30mm){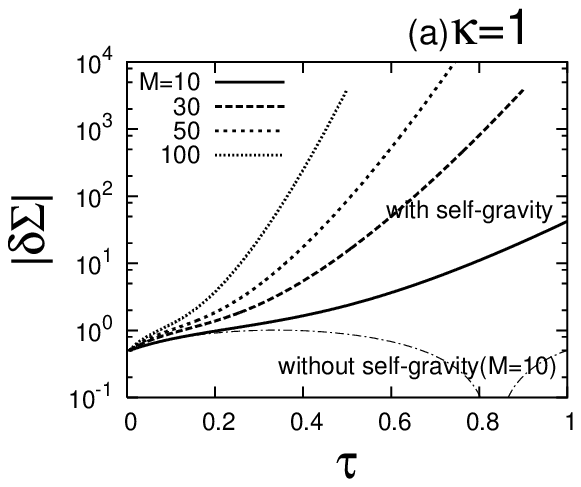} 
     & 
     \FigureFile(80mm,30mm){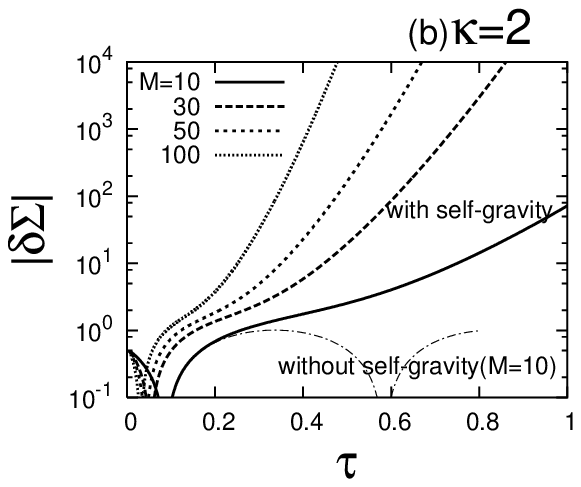} \\
  \end{tabular}
 \end{center}
  \caption{Time evolution of $\delta \Sigma$ for the case with 
           (a)$\kappa=1$ and (b)$\kappa=2$.
           In each panel, 
           thick lines correspond to the case with self-gravity
           (${\cal M\mit}$=10, 30, 50 and 100), 
           while the thin dot-dashed line corresponds to the case
           without self-gravity (${\cal M\mit}=10$).
           }\label{Mach}
\end{figure*}
%%%
\begin{figure*}%[h]
  \begin{tabular}{cc}
   \begin{minipage}{0.5\hsize}
   \centering
  \FigureFile(80mm,30mm){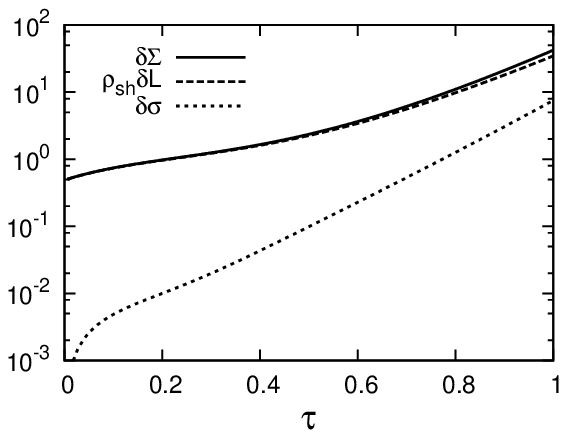}
 \caption{Time evolution of $\delta \Sigma$ (solid line), 
          $\rho_\mathrm{sh}\delta L$ (long-dashed line) and  
          $\delta \sigma$ (short-dashed line) for $\cal M\mit$=10 and $\kappa=1$.
}\label{incompress}
\end{minipage}
 \begin{minipage}{0.5\hsize}
 \centering
  \FigureFile(80mm,30mm){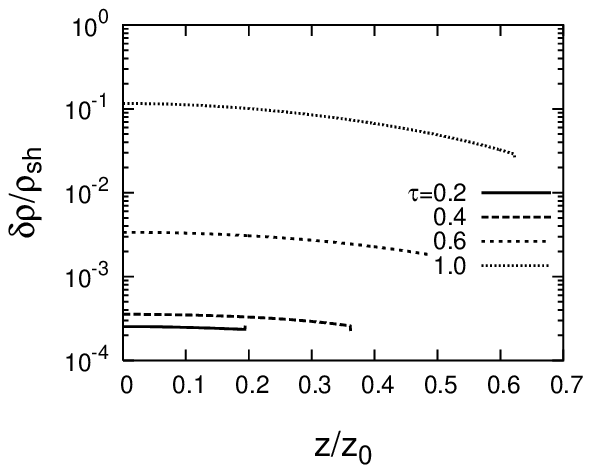}\\
  \caption{Distributions of density perturbation $\delta \rho(z)$ 
          at $\tau=0.2$, 0.4, 0.6 and 1.0.
          The abscissa is normalized by 
          $z_0=c_\mathrm{s}/(\sqrt{2\pi G \rho_\mathrm{E}}{\cal M\mit})$.
          The ordinate is normalized by $\rho_\mathrm{E}{\cal M\mit}^2$.
}\label{den_evolution}
\end{minipage}
\end{tabular}
\end{figure*}
%%%
\begin{figure*}%[h]
 \begin{center}
  \begin{tabular}{cc}
     \FigureFile(90mm,25mm){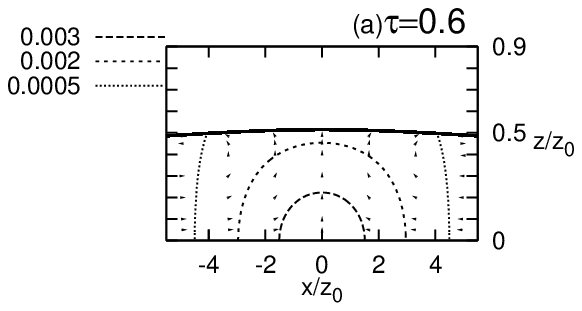} 
     &
     \FigureFile(90mm,25mm){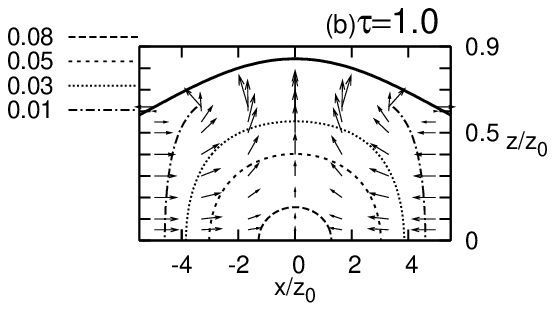} \\
  \end{tabular}
 \end{center}
  \caption{Cross sections of the perturbed layer for ${\cal M\mit}=10$ and $\kappa=2$ at 
           (a)$\tau=0.6$ and (b)$\tau=1.0$. The normalization  
           $\delta z_\mathrm{sh}/z_\mathrm{sh,0}=0.05$ at $\tau=0.6$ is used.
           Contours denote the density perturbation normalized by 
           $\rho_\mathrm{E}{\cal M\mit}^2$, and arrows denote the velocity 
           perturbation fields of which length represents their magnitude.
           The thick solid lines show 
           $z_\mathrm{sh}=z_{\mathrm{sh,0}}+\delta z_\mathrm{sh}$.
           The abscissa and the ordinate  are normalized by 
           $z_0$ as figure \ref{den_evolution}.
           The unperturbed shock position at $\tau=0.6$ and 1.0 
           are $z_{\mathrm{sh,0}}=0.49z_0$ and $0.62z_0$, respectively.}
  \label{structure}
\end{figure*}
%%%
Figure \ref{Mach} shows the time evolution of $\delta \Sigma$
for various Mach numbers, (${\cal M\mit}=10,\;30,\;50$ and $100$). 
Figures \ref{Mach}a and \ref{Mach}b correspond to $\kappa=1$ and 2, respectively.
The thin dot-dashed line shows the result without self-gravity for
${\cal M\mit}=10$. 
The thick solid line shows the result with self-gravity for $\cal M\mit$=10.
In both of figures \ref{Mach}a and \ref{Mach}b,
the lines for ${\cal M\mit}=10$ with and without 
self-gravity are identical initially ($\tau\ltsim 0.1$). Later,
the thick solid line with self-gravity changes to a growing
mode at $\tau\sim 0.2-0.3$.
We define the epoch of transition from an oscillatory mode to a growing mode 
as $\tau_\mathrm{g}$, which corresponds to the inflection point of $\delta \Sigma$ 
($\mathrm{d}^2 \delta \Sigma /\mathrm{d}\tau^2=0)$. 
We will find that $\tau_\mathrm{g}$ is well approximated by 
$t_{c,k}$ in the 1-zone model in subsection \ref{1zone}.
In figures \ref{Mach}a and \ref{Mach}b, it is also seen that 
the growth rate is larger and $\tau_\mathrm{g}$ is smaller for the larger Mach number. 
The dependences of 
the growth rate and $\tau_\mathrm{g}$ on the Mach number are discussed quantitatively in 
subsection \ref{depend}.

As shown in equation (\ref{sigma two contri}), $\delta \Sigma$ consists of 
$\rho_\mathrm{sh}\delta L$ and $\delta \sigma$. In figure \ref{incompress}, 
it can be clearly seen that the fluctuation of the boundary $\rho_\mathrm{sh}\delta L$ 
dominates in the evolution of the perturbations during $\tau\ltsim 1$.
On the other hand, $\delta \sigma$ stays much smaller than $\rho_\mathrm{sh}\delta L$ 
in $\tau\ltsim1$. This means that the layer behaves like an incompressible fluid 
for $\tau\ltsim 1$.
For this reason, as an indicator of the perturbation amplitude, 
we mainly use the column density perturbation, 
$\delta \Sigma$, instead of $\delta \rho(z)$.

Figure \ref{den_evolution} shows the density perturbation, $\delta \rho(z)$,
inside the layer at various epochs, $\tau=0.2$, 0.4, 0.6 and 1.0 for ${\cal M\mit}=10$ 
and $\kappa=1$. 
As normalization, we set $\delta z_{\mathrm{sh}}/z_{\mathrm{sh,0}}=0.05$ 
at $\tau=0.6$. The abscissa and the ordinate axes
are normalized by $z_0=c_\mathrm{s}/(\sqrt{2\pi G \rho_\mathrm{E}}{\cal M\mit})$
and $\rho_\mathrm{sh}={\cal M\mit}^2\rho_\mathrm{E}$, respectively.
The position of the shock front, $z_\mathrm{sh,0}$, is seen as the right edge of 
each line. 
At $\tau \gtsim 0.3$ the perturbations become gravitationally unstable 
(c.f., figure \ref{Mach}a), and $\delta \rho$ increases with time.

%%%
Figure \ref{structure} shows cross sections of the perturbed layer  
at (a)$\tau=0.6$ and (b)$\tau=1.0$. 
The normalization of the perturbation is the same as 
in figure \ref{den_evolution}. 
It can be seen that the perturbation grows gravitationally 
from $\tau=0.6$ to $1.0$. 
Near the equatorial plane, 
inflow in the $x$-direction is generated by gravity, 
which reflects to enhancing the shock front.
In contrast, near the shock boundary, 
the perturbed velocity field is outward in the $x$-direction.
This tangential flow is the result of the shock boundary condition.
This momentum flow acts against the force of gravity (see Appendix \ref{1zone} 
in detail).
%%%
\begin{figure*}[t]
 \begin{center}
  \begin{tabular}{cc}
     \FigureFile(80mm,30mm){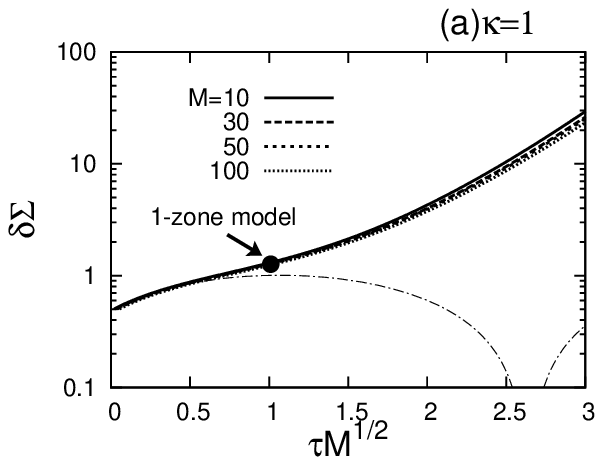} 
     &
     \FigureFile(80mm,30mm){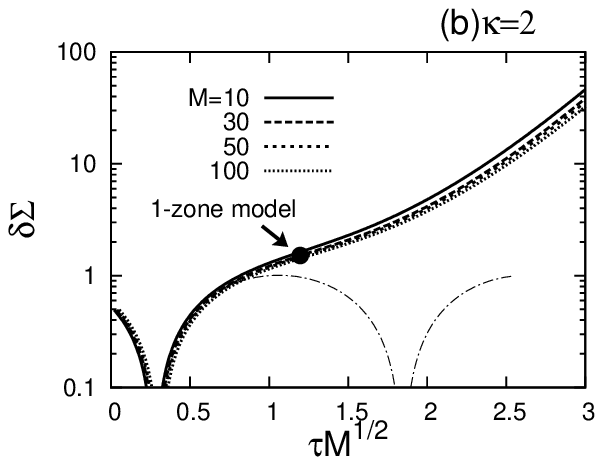}\vspace{2mm}\\
  \end{tabular}
 \end{center}
  \caption{The same figure as the figure \ref{Mach} except that the abscissa denotes normalized time 
           $\tilde{\tau}=\tau {\cal M\mit}$ which is defined in subsection \ref{1zone}.
           Filled circle shows the transition time $\tilde{\tau}_\mathrm{c,\kappa}$ which 
           is predicted from 1-zone model in subsection \ref{1zone}.
           }\label{tsqrtMach}
\end{figure*}
%%%%%%%%%%%%%%%%%
\subsection{The Growth Rate and Transition Epoch}\label{depend}
%%%%%%%%%%%%%%%%%%%
We discuss the dependence of the growth rate and 
the transition epoch, $\tau_\mathrm{g}$, on Mach number. 
In the 1-zone model, according to equation (\ref{tau kappa def}),
the dependence of the transition epoch, $\tau_\mathrm{c,\kappa}$, on Mach number
is ${\cal M\mit}^{-1/2}$.
To see whether the same scaling relation holds or not in the model without 
the 1-zone 
approximation, we replot figure \ref{Mach} in figure \ref{tsqrtMach} using 
$\tilde{\tau}=\tau{\cal M\mit}^{1/2}$ as the abscissa.

In figures \ref{tsqrtMach}a and \ref{tsqrtMach}b,  it is seen that
each line for different Mach numbers is nearly identical.
This result is explained as follows: in the 1-zone model, 
from equation (\ref{disp 1zone}), 
the time evolution of $\delta \Sigma$ is expressed as
%%%
\begin{equation}
\frac{\partial \ln\delta \Sigma}{\partial t}=\sqrt{-\omega^2} 
\;\; \rightarrow \;\;
\frac{\partial \ln\delta \Sigma}{\partial \tilde{\tau}}
=\sqrt{2 \kappa \tilde{\tau} - \kappa^2},
\end{equation}
%%%
where $\tilde{\tau}=\sqrt{\cal M\mit}\tau$. 
Integrating this equation over $\tilde{\tau}$, we obtain
%%%
\begin{equation}
\delta \Sigma \propto
\exp\left[\frac{1}{3\kappa}(2\kappa\tilde{\tau} - \kappa^2)^{3/2}\right].
\label{1zone growth rate}
\end{equation}
%%%
Equation (\ref{1zone growth rate}) shows that the time evolution of 
$\delta \Sigma$ seems to be independent of the Mach number 
if we take $\tilde{\tau}$ as dimensionless time.
Therefore, the growth rate of our calculation is also expected to be independent 
of the Mach number using the normalized time, $t/t_\mathrm{c}$, 
and the wavenumber, $k/k_\mathrm{c}$. 
%%%

Next, we consider the transition epoch from an oscillatory to a growing mode.
In figure \ref{tsqrtMach}, it is clearly seen that $\delta \Sigma$ 
changes from the oscillatory to the growing mode at the same value of $\tau{\cal M\mit}^{1/2}$.
The filled circle indicates the transition epoch $\tilde{\tau}_{c,\kappa}$ predicted 
from the 1-zone model. 
In both of figures \ref{tsqrtMach} (a)$\kappa=1$ and (b)$\kappa=2$, 
it is seen that $\delta \Sigma$ begins to 
grow at $\tau_\mathrm{g}{\cal M\mit}^{1/2}=\tilde{\tau}_\mathrm{c,\kappa}$, 
where $\tau_\mathrm{g}$ 
corresponds to the inflection point of $\delta \Sigma$.
We checked that $\tau_\mathrm{c,\kappa}$ in the 1-zone model gives 
a correct transition epoch even in the
model without the 1-zone approximation, at least for $0.125\leqq\kappa\leqq8$. 

%%%%%%%%%%%%%%%%%
\subsection{Time-Evolving Dispersion Relations}
\begin{figure*}%[h]
 \begin{center}
  \begin{tabular}{cc}
     \FigureFile(80mm,25mm){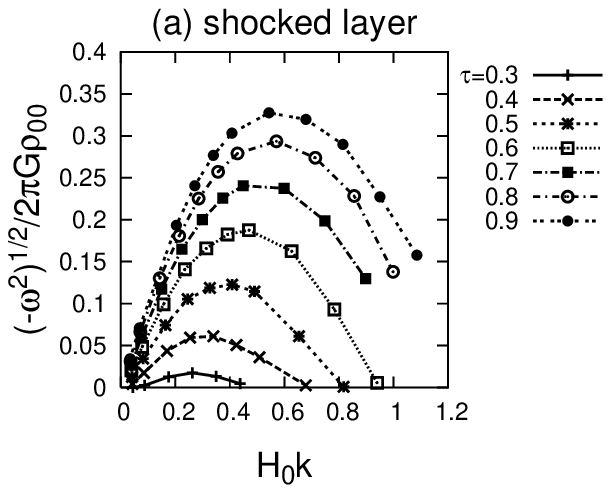} 
     &
     \FigureFile(80mm,25mm){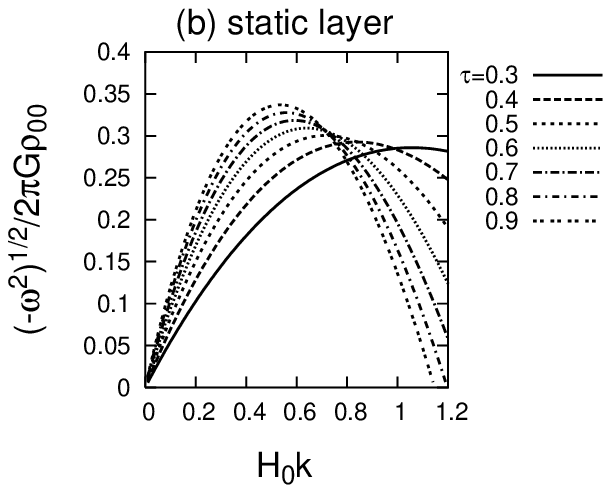}  \vspace{2mm}\\
  \end{tabular}
 \end{center}
  \caption{(a) Dispersion relations derived from our calculation at each epoch for
           ${\cal M\mit}=30$. (b) Dispersion relations of a static isothermal layer
           with the column density and an external pressure at each epoch (see GL65). 
           The abscissa and the ordinate are 
           normalized by the scale height $H_0$ and 
           $2\pi G \rho_{00}$, respectively.
           }\label{omega2}
\end{figure*}

%%%%%%%%%%%%%%%%%%%
Dispersion relations at each instant of time are considered. 
The growth rate at $t$ is defined by 
%%%
\begin{equation}
\omega \equiv \frac{\mathrm{d}}{\mathrm{d}t}
\left\{\ln\left(\frac{\delta \Sigma}{\Sigma_0}\right)\right\}.
\label{omega}
\end{equation}
%%%
Figure \ref{omega2}a shows the dispersion relation derived from 
equation (\ref{omega}) at each instant of time 
for ${\cal M\mit}=30$. Figure \ref{omega2}b shows dispersion relations 
for a static layer corresponding to the density distribution 
at each instant of time (see GL65). 
When $\tau<1$, the distinction between figure \ref{omega2}a and 
figure \ref{omega2}b is obvious. 
In the static layer, 
when $\tau\rightarrow 0$, $(\sqrt{-\omega^2})_\mathrm{max}$ 
and $k_\mathrm{max}^{-1}$ are given by the free-fall growth rate and the 
thickness of the layer, respectively (see Appendix \ref{const}). 
However, in the shocked layer, $(\sqrt{-\omega^2})_\mathrm{max}$ and $k_\mathrm{max}$ 
converge to the zero as $\tau\rightarrow 0$. 
Thus, imposing shock boundary condition is important to understand 
the fragmentation process in the shocked layer. 
\citet{W82} reported a similar property for a layer with shock on one side.
%%%

\begin{figure*}
 \begin{center}
  \begin{tabular}{cc}
     \FigureFile(80mm,25mm){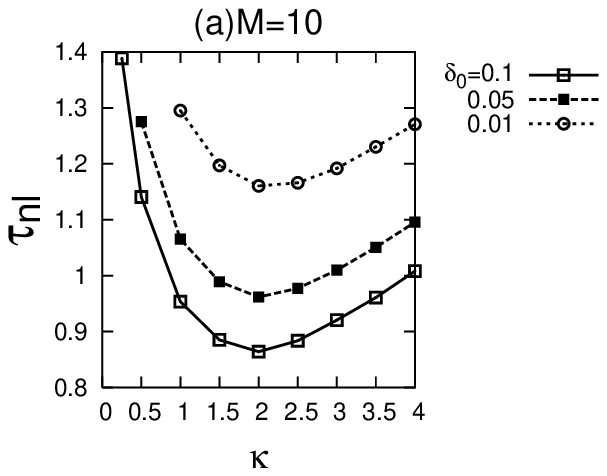}
     &
     \FigureFile(80mm,25mm){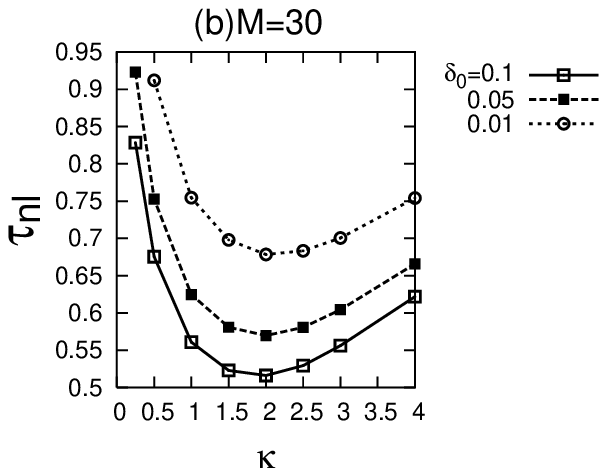} \\
  \end{tabular}
 \end{center}
  \caption{Non-linear time defined by equation (\ref{define tnl}) for
          wavenumbers and initial amplitude for the case with 
           ${\cal M\mit}=$(a)10, (b)30
  }\label{k_t_delta}
\end{figure*}
%%%%%%%%%%%%%%%%%%
\subsection{Fragmentation Time: When Perturbation Becomes Non-linear}
%%%%%%%%%%%%%%%%%
Since the growth rate depends on both the time and the wavenumber 
(figure \ref{omega2}a, figure \ref{begin tau}b), 
the above discussion is not sufficient to predict the fragmentation scale.
Thus, we consider the wavenumber that becomes non-linear at the earliest
epoch. The time $\tau_\mathrm{nl}$ when a mode becomes non-linear is defined by
%%%
\begin{equation}
\frac{\delta \Sigma(\tau_\mathrm{nl})}{\Sigma_0(\tau_\mathrm{nl})}\equiv1.
\end{equation}
%%%
The time $\tau_\mathrm{nl}$ depends on the wavenumber and 
the initial amplitude of the perturbation. 
To focus on the evolution after the transition epoch, $\tau_{c,\kappa}$, 
we define the initial amplitude $\delta_0$ at $\tau_{c,\kappa}$, by
%%%
\begin{equation}
\frac{\delta \Sigma(\tau_{c,\kappa})}{\Sigma_0(\tau_{c,\kappa})} \equiv \delta_0.
\end{equation}
%%%
Combining this definition with $\Sigma_0\propto t$, we obtain
%%%
\begin{equation}
\delta_0\frac{\delta \Sigma(\tau_{\mathrm{nl}})}
{\delta \Sigma(\tau_{\mathrm{c,\kappa}})}
\frac{\tau_{\mathrm{c,\kappa}}}{\tau_{\mathrm{nl}}}=1.
\label{define tnl}
\end{equation}
The arrival time $\tau_\mathrm{nl}$, which satisfies equation 
(\ref{define tnl}), is calculated as a result of our linear analysis.
%%%%%%%%%%%%%%%%%
\subsection{The Fastest Growing mode}\label{fastest mode}
\begin{figure}[h]
 \begin{center}
     \FigureFile(80mm,20mm){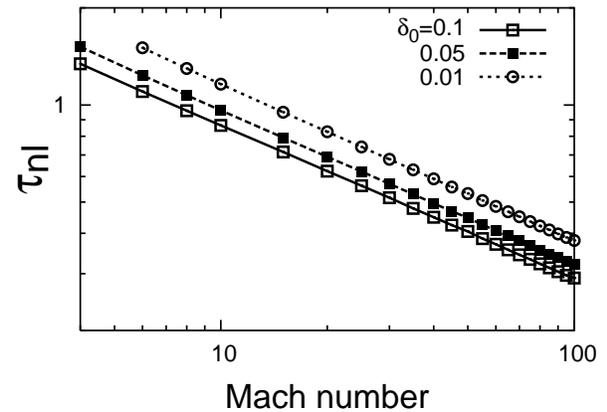} \\
 \end{center}
  \caption{Dependence of $\tau_\mathrm{nl}$ on Mach numbers for the case with $\kappa=2$. 
           Initial amplitude $\delta_0$=0.01, 0.05, and 0.1 are defined 
           in equation (\ref{define tnl})
           }
  \label{Mach t}
\end{figure}

%%%%%%%%%%%%%%%%%
In this section, we determine the wavenumber that becomes non-linear at the 
earliest epoch by using equation (\ref{define tnl}).
Figure \ref{k_t_delta} represents the dependence of $\tau_{\mathrm{nl}}$ on 
the wavenumber for various $\cal M\mit$ and initial amplitude, $\delta_0$.
Figure \ref{k_t_delta} shows that the mode with $\kappa = 2$ becomes
non-linear at the earliest epoch for all ${\cal M\mit}$ and $\delta_0$.
Note that this wavenumber is different from one ($\kappa=1$, figure \ref{begin tau}),
which becomes unstable at the earliest epoch (see subsection \ref{1zone}).
For the fastest growing mode, $\kappa_\mathrm{nl}=2$, the dependence of $\tau_\mathrm{nl}$ 
on the Mach numbers 
is shown in figure \ref{Mach t}. From figure \ref{Mach t}, we can see that 
$\tau_{\mathrm{nl}}\propto {\cal M\mit}^{-1/2}$ holds for each $\delta_0$.
This Mach number dependence is simply understood by the fact that equation 
(\ref{1zone growth rate}) does not contain ${\cal M\mit}$ with normalized 
$t/t_c$ and $k/k_c$.
By fitting the dependence on $\delta_0$, 
$\tau_{\mathrm{nl}}$ is finally given by
%%%
\begin{equation}
\tau_{\mathrm{nl}}\sim 2.4\delta_0^{-0.1}{\cal M\mit}^{-1/2}.
\label{tnl}
\end{equation}
%%%
The error of this fitting formulae is less than 
10\% for $10^{-3}<\delta_0<0.1$. 

Since the dependence on $\delta_0$ is small in equation (\ref{tnl}), if we neglect factors of order unity,
$\tau_\mathrm{nl}$ becomes $\sim {\cal M\mit}^{-1/2}$.
Therefore, $t_\mathrm{nl}=\tau_\mathrm{nl}/\sqrt{2\pi G \rho_\mathrm{E}}\sim 
t_\mathrm{ff}^\mathrm{cloud}/\sqrt{\cal M\mit}$ 
is inversely proportional to the growth rate at $t_\mathrm{c,\kappa}$, where 
$t_\mathrm{ff}^\mathrm{cloud} =1/\sqrt{2\pi G \rho_\mathrm{E}}$.
Note that $t_\mathrm{nl}$ is shorter than the free-fall timescale of the 
preshock region $\sim t_\mathrm{ff}^\mathrm{cloud}$ and 
larger than the free-fall timescale of the layer 
$t_\mathrm{ff}^\mathrm{layer}\sim t_\mathrm{ff}^\mathrm{cloud}/{\cal M\mit}$.
For larger Mach numbers, the difference among
$t_\mathrm{nl}$, $t_\mathrm{ff}
^\mathrm{cloud}$ and $t_\mathrm{ff}^\mathrm{layer}$ becomes large.
The origin of dependence $t_\mathrm{nl}\propto 1/\sqrt{\cal M\mit}$ is explained 
in subsection \ref{1zone}.
In summary,  $\kappa_\mathrm{nl}=2$ and equation (\ref{tnl}) are one of the main results in 
our linear analysis and that provide fragmentation length and time scales.
%%%
%%%%%%%%%%%%%%%%%
\section{Discussion}
%%%%%%%%%%%%%%%%%
\subsection{Astrophysical Implications}
%%%%%%%%%%%%%%%%%
As an application of equation (\ref{tnl}), 
we seek criteria for collision-induced fragmentation.
As a simplest case, 
we consider a spherical cloud with uniform density
$\rho_\mathrm{E}$, the mass $M$ and radius $R$. 
This cloud is specified by one parameter, $\alpha_0$, which is defined by
%%%
\begin{equation}
\alpha_0 = \frac{\cal U\mit_0}{|\cal W\mit_0|}
= \frac{5}{2}\frac{c_\mathrm{s}^2 R}{G M},
\label{def alpha}
\end{equation}
%%%
where $\cal W\mit_0$ and $\cal U\mit_0$ are the gravitational and thermal 
energy of the cloud at the initial state, respectively. 
From equation (\ref{def alpha}) and $M\simeq 4\pi \rho_\mathrm{E} R^3/3$, 
the radius is given by 
%%%
\begin{equation}
2R = \sqrt{\frac{15 c_\mathrm{s}^2}{2\pi G \rho_\mathrm{E} \alpha_0}}.
\label{radius}
\end{equation}
%%%
We consider the collision of two physically identical clouds with velocity 
$\pm{\cal M\mit}c_{\mathrm{s}}$ relative to the center of mass. 

Fragmentation during cloud collision is expected if the following two
conditions are satisfied (hereafter we adopt $t_\mathrm{nl}$ as
fragmentation time).
%%%%%%%%%%%%%%%%%%%%%%%%%%%%%%%%%%
\begin{enumerate}
\item \mbox{\boldmath $\tau_{\mathrm{nl}} \ltsim \tau_{\mathrm{conti}}$ }
%%%%%%%%%%%%%%%%%%%%%%%%%%%%%%%%%%
The first condition is that fragmentation time, $\tau_\mathrm{nl}$, is 
shorter than the normalized collision continuous time, $\tau_{\mathrm{conti}}$,
which is defined by
%%%
\begin{equation}
\tau_{\mathrm{conti}}= \frac{2R}{{\cal M\mit}c_{\mathrm{s}}}
\sqrt{2\pi G \rho_\mathrm{E}}.
\label{conti time}
\end{equation}
%%%
From equation (\ref{radius}) and (\ref{conti time}), we obtain
%%%
\begin{equation}
\tau_{\mathrm{conti}} \simeq \sqrt{15}\alpha_0^{-1/2}{\cal M\mit}^{-1}.
\label{conti time 1}
\end{equation}
%%%
Substituting equation (\ref{conti time 1}) and 
our result of linear analysis [equation (\ref{tnl})]
into $\tau_{\mathrm{nl}}\ltsim \tau_{\mathrm{conti}}$, the first condition 
for fragmentation is written as 
%%%
\begin{equation}
\alpha_0 \lesssim 2.6\delta_0^{0.2}{\cal M\mit}^{-1}.
\label{cond1}
\end{equation}
%%%%%%%%%%%%%%%%%%%%%%%%%%%%%%%%%%
\item \mbox{\boldmath $2c_\mathrm{s} t_\mathrm{conti} \ltsim 2R - \lambda_\mathrm{nl}$ }
%%%%%%%%%%%%%%%%%%%%%%%%%%%%%%%%%%
In order to assume the compressed region between two clouds as a layer,
rarefaction wave from the cloud surface must not erode in an unstable 
region with $\lambda_\mathrm{nl}$, which is given by
\begin{equation}
\lambda_\mathrm{nl}=\frac{2\pi}{k_\mathrm{nl}}\simeq 
2\pi \frac{c_\mathrm{s}}{2\sqrt{2\pi G \rho_\mathrm{E}{\cal M\mit}}}.
\label{lam nonli}
\end{equation}

This condition is expressed as $2c_\mathrm{s}t_\mathrm{conti} 
\ltsim 2R - \lambda_\mathrm{nl}$. With equation (\ref{radius}), 
(\ref{conti time 1}) and (\ref{lam nonli}), the above condition is  rewritten as
%%%
\begin{equation}
\alpha_0 \ltsim \frac{15}{\pi^2}{\cal M\mit}\left(1-\frac{2}{\cal M\mit}\right)^2.
\label{cond2}
\end{equation}
%%%
\end{enumerate}
%%%
Therefore, two conditions, (\ref{cond1}) and (\ref{cond2}),
provide fragmentation criteria on $\alpha_0$ and $\cal M\mit$.

\begin{figure}
 \begin{center}
     \FigureFile(80mm,30mm){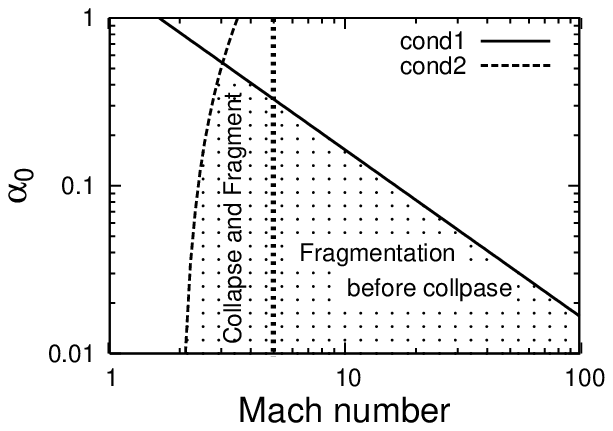} \\
 \end{center}
  \caption{Criterion of fragmentation. 
           The solid and long-dashed lines correspond to 
           $\tau_{\mathrm{nl}}=\tau_{\mathrm{conti}}$ and 
           $2c_\mathrm{s} t_\mathrm{conti} = 2R - \lambda_\mathrm{nl}$, respectively.
           The fragmentation region is shown to be in the dotted region.
           The short-dashed line represents
           $\tau_\mathrm{nl}=\tau_\mathrm{dyn}({\cal M\mit}={\cal M\mit}_\mathrm{eq})$.
           The region with
           $\tau_\mathrm{nl}>\tau_\mathrm{dyn}({\cal M\mit}<{\cal M\mit}_\mathrm{eq})$
           shows that the layer probably fragments during individual collapse.
           The region with
           $\tau_\mathrm{nl}<\tau_\mathrm{dyn}({\cal M\mit}>{\cal M\mit}_\mathrm{eq})$
           shows that the layer can fragment before individual collapse.
           }\label{alpha Mach}
\end{figure}

Figure \ref{alpha Mach} represents the two conditions on the 
($\alpha_0,{\cal M\mit}$) plane.
Fragmentation is expected in the dotted region.
Figure \ref{alpha Mach} indicates that the case with $\alpha_0\simeq 1$ 
can not be expected to fragment in all Mach numbers. 
This means that the collision of clouds in dynamical equilibrium 
does not induce fragmentation.
By collision between clouds, only the case with small $\alpha_0$
is expected to fragment the layer. In other words,  
only clouds with mass much larger than Jeans mass will fragment by cloud 
collision as long as isothermal. 

If $\alpha_0\ltsim1$, each cloud collapses due to self-gravity. 
The collapse timescale is 
$t_\mathrm{dyn}=\sqrt{3\pi/(32 G \rho_\mathrm{E})}$, 
or $\tau_\mathrm{dyn}= \sqrt{3}\pi/4$.
The critical Mach number, $\cal M\mit_\mathrm{eq}$,
in which $\tau_\mathrm{nl}$ equals to $\tau_\mathrm{dyn}$,
is given by
%%%
\begin{equation}
{\cal M\mit}_\mathrm{eq} \sim  3.1\delta_0^{-0.2}.
\label{cond3}
\end{equation}
%%%
In Figure \ref{alpha Mach}, thick dotted vertical line represents 
${\cal M\mit}={\cal M\mit}_\mathrm{eq}$.
In the region with ${\cal M\mit}\ltsim{\cal M\mit}_\mathrm{eq} 
\;(t_\mathrm{nl}\gtsim t_\mathrm{dyn})$, 
the dynamical collapse of each cloud can not be ignored during fragmentation.
Since our linear analysis does not include collapse of each cloud,
a non-linear multi-dimensional calculation is required for further an investigation.
In the region with ${\cal M\mit}\gtsim{\cal M\mit}_\mathrm{eq}
\;(t_\mathrm{nl}\ltsim t_\mathrm{dyn})$,  
the layer is expected to fragment before an individual collapse.

\citet{NM87} investigated collisions between self-gravitating isothermal equilibrium
spheres supported by an external pressure using three-dimensional calculations. 
They calculated cases by coalescence between two clouds with $\alpha_0 \sim o(1)$ 
and ${\cal M\mit}\lesssim10$.
As a result, they found that a single cloud forms and it
becomes gravitationally unstable. However, it does not fragment, 
but collapses as a single object. Our results are consistent with their results 
in $\alpha_0 \sim o(1)$ and ${\cal M\mit}\lesssim10$.
However, in order to compare our results and to discuss the fragmentation 
process of the layer which is formed by a collision between clouds with $\alpha_0<1$,
a much higher numerical resolution than previous calculation is required.

\citet{BW05} investigated the evolution of an oblate spheroidal cloud formed by a shocked layer 
using a semi-analytic method. 
Their clouds correspond to the unit wavelength in out model, while they included the effect of 
detailed thermal evolution.
A cloud smaller than $\sim c_s/\sqrt{2\pi G \rho_\mathrm{E}{\cal M\mit}}$ oscillate, and then
collapse, while larger clouds collapse monotonically.
Their results are consistent with ours.

Finally, we comment on two possibilities of fragmentation that
is induced by cloud collision in an actual astrophysical environment.
The first possibility is a collision between clouds with $\alpha_0 < 1\;(M>M_\mathrm{J})$.
From figure \ref{alpha Mach}, the timescale of collision-induced fragmentation 
is smaller than that of the collapse of each cloud if collision velocity 
is ${\cal M\mit}\gtsim{\cal M\mit}_\mathrm{eq}\sim 5$ [equation (\ref{cond3})]. 
Therefore sheet fragmentation becomes dominant process in such a case of cloud collision.
The second possibility is that $\alpha_0$ becomes effectively smaller than order of unity 
after formation of the layer. 
This will be realized if the postshock region becomes cooler than the preshock region by cooling,
even in a collision between clouds with $\alpha_0 \sim 1$.
%%%
%%%%%%%%%%%%%%%%%%%%%%%%%%%%%%%%%%%%%%%%%%%%%%%%%%%%%%%%%%%%%%%%%%%%%%%%%%%%%%
\section{Summary}
%%%%%%%%%%%%%%%%%%%%%%%%%%%%%%%%%%%%%%%%%%%%%%%%%%%%%%%%%%%%%%%%%%%%%%%%%%%%%
We investigated the gravitational instability of isothermal layers bounded 
by shock waves. 
The results of our investigation are summarized as follows. 
%%%
\begin{enumerate}
%%%
\item
By imposing the shock boundary condition, the shocked layer is more stabilized 
than static layer, which is bounded by a constant external pressure.
The stabilization influences the fragmentation process of the shocked layer.
\vspace{1mm}
%%%
\item
In our linear analysis, the epoch at which the perturbations change 
from the oscillatory mode
to the growing mode is well approximated by $t_{c,\kappa}$ [see equation (\ref{tau c})] 
derived from the 1-zone model.
\vspace{1mm}
%%%
\item
Our unsteady linear analysis can provide the correct fastest 
growing mode, which becomes non-linear at the earliest epoch.
The wavenumber of the fastest growing mode is
$k_\mathrm{nl}=2\sqrt{2\pi G \rho_\mathrm{E}{\cal M\mit}}/c_\mathrm{s}$ 
($\kappa_\mathrm{nl}=2$). 
This value, $\kappa_\mathrm{nl}=2$, is twice as large as
the wavenumber, $\kappa=1$, which begins to grow at the earliest epoch. 
The time at which this mode becomes non-linear is given by $t_\mathrm{nl}
= 2.4\delta_0^{-0.1}/\sqrt{2\pi G \rho_\mathrm{E}{\cal M\mit}}$.
\vspace{1mm}
\item
Collision-induced fragmentation is expected only when
$\alpha_0=5c_\mathrm{s}^2 R/2G M$ of the parent clouds is much smaller than 1 
(figure \ref{alpha Mach}).
In the collision between clouds with $\alpha_0<1$ and ${\cal M\mit}\gtsim 5$, 
the layer will fragment before individual collapse by self-gravity.
%Even in collision between clouds with $\alpha_0\sim 1$, 
%the layer may fragment during collision 
%if postshock region becomes cooler than preshock region by cooling and 
%$\alpha_0$ effectively becomes $<1$ effectively.
\end{enumerate}
%%%%%%%%%%%%%%%%%%%%%%%%%%%%%%%%%%%%%%%%%%%%%%%%%%%%%%%%%%%%
\section*{Acknowledgements}
%%%%%%%%%%%%%%%%%%%%%%%%%%%%%%%%%%%%%%%%%%%%%%%%%%%%%%%%%%%%
We thank Fumio Takahara and Yutaka Fujita for useful discussion and continuous encouragement.
This work is in part supported by the 21st Century COE Program 
"Towards a New Basic Science; Depth and Synthesis" in Osaka University, 
funded by the Ministry of Education, Science, Sports and Culture of Japan.
%%%%%%%%%%%%%%%%%%%%%%%%%%%%%%%%%%%%%%%%%%%%%%%%%%%%%%%%%%%%
\appendix
\section{The Shock Boundary Conditions}\label{shock boundary}
%%%%%%%%%%%%%%%%%%%%%%%%%%%%%%%%%%%%%%%%%%%%%%%%%%%%%%%%%%%%
%%%%%%%%%%%%%%%%%
\subsection{Unperturbed Shock Boundary Conditions}
%%%%%%%%%%%%%%%%%
\begin{figure}[h]
 \begin{center}
     \FigureFile(80mm,30mm){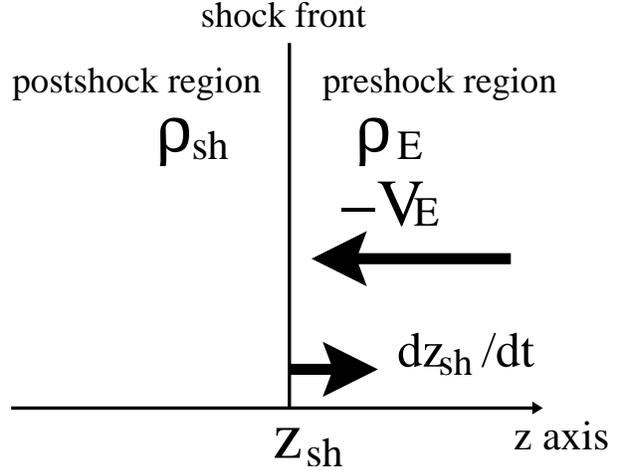} 
 \end{center}
 \caption{Schematic picture of shock front in the rest frame of postshock.}
\label{rankine.down}
\end{figure}
In the rest frame of the postshock, we derive jump conditions of isothermal shock
without self-gravity.
A schematic picture is represented in figure \ref{rankine.down}. 
The physical variables in the preshock and the postshock regions are denoted
by subscripts of "E" and "$\mathrm{sh}$", respectively.

To begin, jump conditions in the unperturbed state are considered.
From mass and momentum conservation across the shock front and the 
isothermal condition, we can derive the following equations:
%%%
\begin{equation}
\rho_{\mathrm{sh}}(-\dot{z}_{\mathrm{sh}}) = 
\rho_\mathrm{E} (-V_\mathrm{E} - \dot{z}_{\mathrm{sh}}),
\label{shock cond mass}
\end{equation}
%%%
\begin{equation}
P_{\mathrm{sh}}+ \rho_{\mathrm{sh}}(-\dot{z}_{\mathrm{sh}})^2 =
P_\mathrm{E} + \rho_\mathrm{E} (-V_\mathrm{E} - \dot{z}_{\mathrm{sh}})^2,
\label{shock cond momentum}
\end{equation}
%%%
and
%%%
\begin{equation}
\frac{P_{\mathrm{sh}}}{\rho_{\mathrm{sh}}} 
= \frac{P_\mathrm{E}}{\rho_\mathrm{E}}=c_\mathrm{s}^2.
\label{shock cond isothermal}
\end{equation}
%%%
We define the flow velocity $V_\mathrm{sh}$ in the rest frame of the shock front by 
$V_{\mathrm{sh}} = V_\mathrm{E} + \dot{z}_{\mathrm{sh}}$.
Hence, by solving equations (\ref{shock cond mass}), (\ref{shock cond momentum}) and 
(\ref{shock cond isothermal}), we obtain 
%%%
\begin{equation}
\rho_{\mathrm{sh}}=\rho_\mathrm{E} {\cal M\mit}^2,
\end{equation}
%%%
\begin{equation}
\dot{z}_{\mathrm{sh}}=\frac{c_\mathrm{s}}{\cal M\mit},
\end{equation}
%%%
and 
%%%
\begin{equation}
V_\mathrm{E} = \left({\cal M\mit} - \frac{1}{\cal M\mit}\right)c_\mathrm{s}.
\end{equation}
%%%%%%%%%%%%%%%%
\subsection{Perturbed Shock Boundary Conditions}
%%%%%%%%%%%%%%%%
Linearized jump conditions are derived.
The position of the shock front is defined 
by $z_{\mathrm{sh}}(x,t)=z_{\mathrm{sh,0}}(t)
+ \delta z_{\mathrm{sh}}(t)e^{ikx}$.
Linearized mass and momentum conservation across the shock front 
are given by
%%%
\begin{equation}
D \rho_{\mathrm{sh}}(-\dot{z}_{\mathrm{sh}})
+ \rho_{\mathrm{sh,0}}(D v_{z\mathrm{sh}} - \delta \dot{z}_{\mathrm{sh}})
= -\rho_\mathrm{E} \delta \dot{z}_{\mathrm{sh}},
\label{perturb den shock}
\end{equation}
%%%
and 
%%%
\vspace{-2mm}
\begin{eqnarray}
\lefteqn{D \rho_{\mathrm{sh}}(c_\mathrm{s}^2 + \dot{z}_{\mathrm{sh}}^2)
- 2\rho_{\mathrm{sh,0}} \dot{z}_{\mathrm{sh}}(D v_{z,\mathrm{sh}}
- \delta \dot{z}_{\mathrm{sh}})} \hspace{50mm} \nonumber \\
&=& 2\rho_\mathrm{E} V_{\mathrm{sh}}\delta \dot{z}_{\mathrm{sh}},
\label{perturb vz shock}
\end{eqnarray}
%%%
where operator "$D$" represents Lagrangian displacement.
From equations (\ref{perturb den shock}) and (\ref{perturb vz shock}), 
we have
%%%
\begin{equation}
D\rho_{\mathrm{sh}} = 2\rho_\mathrm{E}\frac{\cal M\mit}{c_\mathrm{s}}
\frac{\mathrm{d}\delta z_{\mathrm{sh}}}{\mathrm{d} t},
\end{equation}
%%%
and
%%%
\begin{equation}
Dv_{z,\mathrm{sh}} = \left(1+\frac{1}{{\cal M\mit}^2}\right)
\frac{\mathrm{d}\delta z_{\mathrm{sh}}}{\mathrm{d} t}.
\end{equation}
%%%
The relationship between the Lagrangian and Eulerian displacement is given
by
%%%
\begin{equation}
D\rho_{\mathrm{sh}} = D\rho_{\mathrm{sh}}(z_{\mathrm{sh,0}}
+ \delta z_{\mathrm{sh}}) = \delta \rho_{\mathrm{sh}}
+ \delta z_{\mathrm{sh}}\left(\frac{\partial \rho_0}{\partial z}\right)
_{z_{\mathrm{sh,0}}}.
\end{equation}
%%%
Next, we consider the velocity perturbation in the $x$-direction
$\delta v_{x,\mathrm{sh}}$. 
Momentum conservation for the tangential direction of the shock front gives
%%%
\begin{equation}
\delta v_{x,\mathrm{sh}} = -V_\mathrm{E} 
\frac{\partial \delta z_{\mathrm{sh}}}{\partial x}
= -ikV_\mathrm{E}\delta z_{\mathrm{sh}}.
\end{equation}
%%%
We adopt the same boundary condition of the gravitational potential perturbation 
as in \citet{W82},
%%%
\begin{equation}
\frac{\partial \delta \phi_\mathrm{sh}}{\partial z}+k\delta \phi_\mathrm{sh} 
+ 4\pi G (\rho_\mathrm{sh} - \rho_\mathrm{E}) =0.
\end{equation}
%%%

%%%%%%%%%%%%%%%%%%%%%%%%%
\section{Dispersion Relation of a Shocked Layer under 1-zone Model}\label{appe 1zone}
%%%%%%%%%%%%%%%%%%%%%%%%%
The averaged physical quantities are defined by 
%%%
\begin{equation}
\Sigma = \int_{-z_\mathrm{sh}}^{z_\mathrm{sh}}\rho\mathrm{d}z\;\;\mathrm{and}\;\;
J_x = \int_{-z_\mathrm{sh}}^{z_\mathrm{sh}}\rho v_x\mathrm{d}z.
\end{equation}
%%%
The 0th-order equation is given by
%%%
\begin{equation}
\partial_t \Sigma_0 = \rho_\mathrm{E} V_{\mathrm{sh}},
\end{equation}
%%%
where the subscript "0" represents the unperturbed variable
and differential operators
$\partial/\partial t$ and $\partial/\partial x$ are abbreviated 
by $\partial_t$ and $\partial_x$, respectively.
Let us consider the perturbation variables $\delta \Sigma$, 
$\delta J_x$ and $\delta L$, which are defined by 
%%%
\begin{equation}
\Sigma(x,t) = \Sigma_0(t) + \delta \Sigma(x,t),
\end{equation}
%%%
\begin{equation}
J_x(x,t) = \delta J_x(x,t),
\end{equation}
%%%
and
%%%
\begin{equation}
L(x,t) = L_0(t) + \delta L(x,t)
,\; L(x,t) = 2 z_{\mathrm{sh}}(x,t).
\end{equation}
%%%
The column density perturbation, $\delta \Sigma$, can be divided into 
two contributions, as follows: 
%%%
\begin{equation}
\delta \Sigma = \rho_{\mathrm{sh}}\delta L + \delta \sigma,
\;\; \mathrm{where}\;\;\delta \sigma =\int_{-z_{\mathrm{sh,0}}}^{z_{\mathrm{sh,0}}} 
\delta \rho \mathrm{d}z .
\label{two contri}
\end{equation}
%%%%

The linearlized equation of mass conservation is given by
%%%%
\begin{equation}
\partial_t \delta \Sigma = \rho_\mathrm{E}\partial_t \delta L 
- \partial_x \delta J_x \sim -\partial_x \delta J_x.
\label{Sigma 1zone nograv}
\end{equation}
%%%%
Since $\delta \Sigma \sim \rho_\mathrm{sh}\delta L \gg \rho_\mathrm{E}\delta L$,
linearlized equation of momentum conservation in the $x$-direction is given by
%%%%
\begin{eqnarray}
\partial_t \delta J_x &=& \rho_\mathrm{E}V_\mathrm{sh}(-V_\mathrm{E}\partial_x \delta L)
 -c_\mathrm{s}^2\partial_x \delta \sigma +2\pi G\Sigma_0\partial_x \delta \Sigma
/k \nonumber \\
&\sim& -\rho_\mathrm{E}{\cal M\mit}^2 \partial_x \delta L 
-c_\mathrm{s}^2 \partial_x \delta \sigma + 2\pi G\Sigma_0\partial_x \delta \Sigma/k. 
\nonumber \\
&=& -c_\mathrm{s}^2 \partial_x \delta \Sigma + 2\pi G\Sigma_0\partial_x \delta \Sigma/k.
\label{J_x 1zone grav}
\end{eqnarray}
%%%%
The second term in the first line of the equation (\ref{J_x 1zone grav}) represents  
the momentum flux as a result of the shock boundary condition.
The second line of the equation (\ref{J_x 1zone grav}) is derived using
$V_\mathrm{E}\sim {\cal M\mit}c_\mathrm{s}$. 
The third line is derived
by using $\rho_\mathrm{sh}=\rho_\mathrm{E}{\cal M\mit}^2$ and equation (\ref{two contri}).

Neglecting the time-dependence of $\Sigma_0(t)$, 
from equation (\ref{Sigma 1zone nograv}) and (\ref{J_x 1zone grav}),
the approximate dispersion relation is derived as
%%%%
\begin{equation}
\omega^2 \sim c_\mathrm{s}^2 k^2 - 2\pi G k \Sigma_0(t).
\label{disp app}
\end{equation}
%%%%
%%%%%%%%%%%%%%%%%%%%%%%%%
\section{A Layer Confined by Constant External Pressure}\label{const}
%%%%%%%%%%%%%%%%%%%%%%%%%
We investigate a highly compressed static layer that is mainly 
confined by constant external pressure,
$P_\mathrm{b}$. In this situation, the layer 
has almost uniform density, denoted by $\rho_0$.
The momentum conservation equation in $x$-direction is given by
%%%
\begin{equation}
\partial_t \delta J_x = 
- c_\mathrm{s}^2 \partial_x \delta \sigma + 2\pi G\Sigma_0 \partial_x \delta \Sigma/k,
\label{J_x const}
\end{equation}
%%%
where the second term of first line of equation (\ref{J_x 1zone grav}) is absent 
due to the constant pressure boundary conditions.
Because the layer is very thin, 
$\delta \rho$ is approximately determined only by boundary value of $\delta \rho$ as
$\delta \sigma \simeq \delta \rho(z=z_{\mathrm{b}}) L_0$, where $z_\mathrm{b}$ is 
the position at the boundary of the layer.
The constant pressure boundary condition is given by
%%%
\begin{equation}
D\rho=\delta \rho(z_\mathrm{b,0}) 
+ \delta z_\mathrm{b}(\partial_z \rho_0)_{z_\mathrm{b,0}}=0,
\end{equation}
%%%
(see GL65). Because the unperturbed state is hydrostatic equilibrium,
$(\partial_z \rho_0)_{z_\mathrm{b}}=
-\rho_0 (\partial_z \phi_0)_{z_\mathrm{b}}/c_\mathrm{s}^2
= -2\pi G \rho_0^2 L_0/c_\mathrm{s}^2$. 
Therefore, the ratio of $\delta \sigma$ to $\rho_0\delta L$ is 
written as
%%%
\begin{equation}
\delta \sigma /\rho_0\delta L\sim (L_0/H_0)^2 \ll 1.
\label{ratio sigma L}
\end{equation}
%%%
From $\delta \Sigma \sim \rho_0\delta L$ and equation (\ref{ratio sigma L}), 
the second term of the equation (\ref{J_x const}) becomes
%%%
\begin{equation}
-c_\mathrm{s}^2 \partial_x \delta \sigma 
\sim - c_\mathrm{s}^2\left(\frac{L_0}{H_0}\right)^2 \partial_x \delta \Sigma.
\end{equation}
%%%
Thus, the pressure term is $\sim (L_0/H_0)^2$ times
as small as that of the shocked layer [equation (\ref{disp 1zone})].
Moreover, $c_\mathrm{s}(L_0/H_0)$ is rewritten with equation (\ref{equilibrium}) as 
%%%
\begin{equation}
c_\mathrm{s}\left(\frac{L_0}{H_0}\right)\sim \sqrt{g_0 L_0},\hspace{2mm} 
g_0\equiv 2\pi G \rho_{00}L_0 \sim 2\pi G \Sigma_0,
\end{equation}
%%%
where $\Sigma_0 \simeq \rho_{00}L_0$.
This corresponds to the phase velocity of gravity wave in shallow water.
Finally, the dispersion relation is given by 
%%%
\begin{equation}
\omega^2 \sim \left(\frac{L_0}{H_0}\right)^2 
c_\mathrm{s}^2 k^2 - 2\pi G k \Sigma_0
\sim 2\pi G \Sigma_0(L_0k^2 - k)
\label{disp 1zone const}
\end{equation}
%%%
From equation (\ref{disp 1zone const}), $(\sqrt{-\omega^2})_\mathrm{max}$ is 
almost the free-fall growth rate of the layer and $\lambda_\mathrm{max}$ is nearly 
equal to the thickness $\sim L_0$. 
The smaller scale perturbation can grow much faster than in the shocked layer 
because the pressure effect is small. 

From the above simple analysis, we understand the following things.
For $(L_0/H_0)\simeq 1$, the difference of boundary condition does not influence on 
the evolution of the perturbations, and the dispersion relations 
become the same form as equation (\ref{disp app}).
However, for $(L_0/H_0)^2\ll 1$, the constant-pressure boundary condition is in 
completely different from that of the shock boundary condition.
%%%%%%%%%%%%%%%%%%%%%%%%%%%%%%%%%%%%%%%%%%%%%%%%%%%%%%%%%%%%%%%%%%%%
\section{Validity of Approximation($v_{z,0}=0$)}\label{valid}
\begin{figure*}%[t]
 \begin{center}
  \begin{tabular}{cc}
     \FigureFile(80mm,30mm){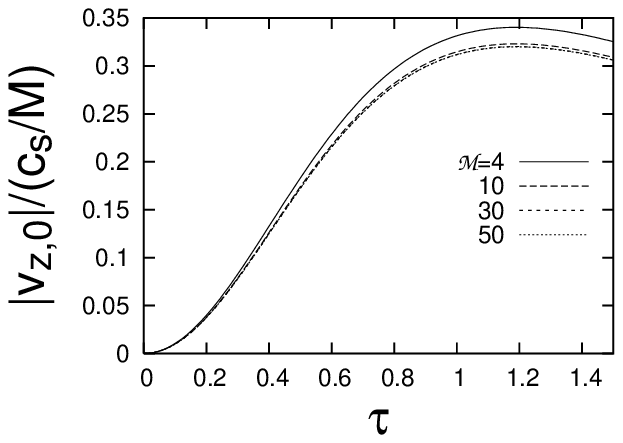} 
     &
     \FigureFile(80mm,30mm){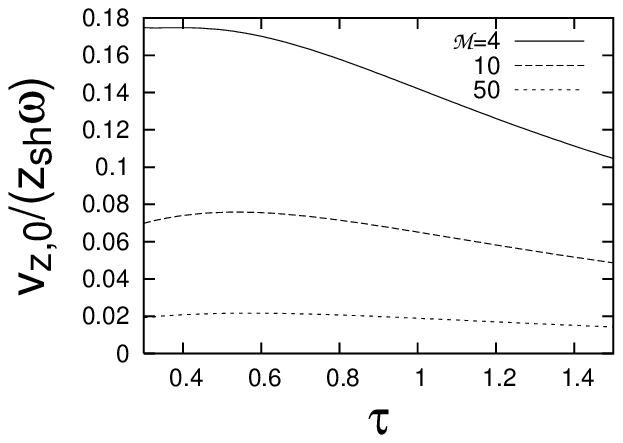} \\
     {\scriptsize (a)} 
     &
     {\scriptsize (b)}\\
  \end{tabular}
 \end{center}
  \caption{(a): Analytic expression (\ref{vz0}) is shown for
           ${\cal M\mit}=4$, 10, 30, 50. 
           (b): Plots of $v_{z,0}/(z_\mathrm{sh}\omega)$. Each variable is from equation 
            (\ref{vz0})($v_{z,0}$), 
           equation (\ref{disp 1zone})($\omega$) and 
           equation (\ref{zsh})($z_\mathrm{sh}$). 
           }\label{neg vz}
\end{figure*}
%%%%%%%%%%%%%%%%%%%%%%%%%%%%%%%%%%%%%%%%%%%%%%%%%%%%%%%%%%%%%%%%%%%%
The validity of the approximation $v_{z,0}=0$ is considered in this section.
The perturbation equations containing $v_{z,0}$ are given by
%%%
\begin{equation}
\frac{\partial \delta \rho}{\partial t}
+ \frac{\partial}{\partial z}(\rho_0 \delta v_z + \delta \rho v_{z,0})
= -ik \rho_0 \delta v_x,
\label{perturb rho}
\end{equation}
%%%
\begin{eqnarray}
\lefteqn{\frac{\partial}{\partial t}(\rho_0\delta v_z + \delta \rho v_{z,0})
+ \frac{\partial}{\partial z}\left(c_\mathrm{s}^2\delta \rho 
+ 2\rho_0 v_{z,0} \delta v_z +
\delta \rho v_{z,0}^2 \right)} \hspace{20mm} \nonumber \\
& =& - ik \rho_0 v_{z,0}\delta v_x - \delta \rho \frac{\partial \phi_0}{\partial z}
- \rho_0 \frac{\partial \delta \phi}{\partial z},
\label{perturb momz}
\end{eqnarray}
%%%
and 
%%%
\begin{equation}
\frac{\partial}{\partial t}(\rho_0\delta v_x)
+ \frac{\partial}{\partial z}(\rho_0 v_{z,0}\delta v_x)
= -ik c_\mathrm{s}^2 \delta \rho - ik\rho_0 \delta \phi.
\label{perturb momx}
\end{equation}
%%%
For $|z|<|z_\mathrm{sh}|$, the velocity $|v_{z,0}(z)|$ is smaller 
than $|v_{z,0}(z_{\mathrm{sh}})|$ at shock front.
If we assume that the shock front position is given by equation (\ref{zsh}), 
we can express $v_{z,0}(z_\mathrm{sh})$ approximately from the jump conditions
%%%
\begin{equation}
v_{z,0}(z_{\mathrm{sh}})=
-\frac{c_\mathrm{s}}{\cal M\mit}
\left(\frac{1}{1-f(\tau)/{\cal M\mit}^2} + f(\tau) - 1 \right),
\label{vz0}
\end{equation}
%%%
where $f$ is given by 
%%%
\begin{equation}
f(\tau)=\tau \cosh^{-1}\sqrt{1+\tau^2}/(1+\tau)^{3/2}.
\end{equation}
%%%
Figure \ref{neg vz}a shows the time evolution of $|v_{z,0}(z_{\mathrm{sh}})|$
for various Mach numbers.
This figure indicates that the maximum value of $|v_{z,0}(z_{\mathrm{sh}})|$
is slightly less than, or comparable to, $\sim 0.3 c_\mathrm{s}/{\cal M\mit}$. 
The average velocity, $|\bar{v}_{z,0}|$, of the layer 
is clearly smaller than this value.

Let us consider the contribution of $v_{z,0}(z_{\mathrm{sh}})$
for the perturbation equations.
To begin with, we compare the two terms $\delta \rho v_{z,0}$ and $\rho_0 \delta v_z$ 
in perturbed momentum in the $z$-direction. The ratio of the two terms
is given by $(\delta \rho v_{z,0})/(\rho \delta v_{z,0}) 
< 0.3c_\mathrm{s}(\delta \rho/\delta v_{z,0})/{\cal M\mit}^3 \ll 1$; thus,
we can approximate the perturbed momentum in the $z$-direction as $\rho_0 \delta v_z$. 
Therefore, we can neglect the third term of equation (\ref{perturb rho}) and 
the second term of equation (\ref{perturb momz}).
Moreover, we can neglect the fifth term of equation (\ref{perturb momz}),
which is smaller than the third term by $0.3/{\cal M\mit}^2$.
In equation (\ref{perturb momz}), the ratio of the fourth to the first terms 
is given by $\sim v_{z,0}/(L_0\omega)$. This ratio 
is small value if Mach number is large (see figure \ref{neg vz}b).
For the same reason, we can neglect the second term of 
equation (\ref{perturb momx}) compared with the first term.
The ratio of the sixth term of equation (\ref{perturb momz}) 
to the first term is given by
$(k/\omega)v_{z,0}(\delta v_x/\delta v_z) \lesssim 
(0.3/{\cal M\mit})(\delta v_x/\delta v_z)$. 
This value is also small if the Mach number is large.
From the above discussion, $v_{z,0}$ can be assume to be zero.

%%%%%%%%%%%%%%%%%%%%%%%%%%%%%%%%%%%%%%%%%%%%%%%%%%%%%%%%%%%%%%%%%%%%%%%%%%%%%%%%

\end{document}